# How Many Ore-Bearing Asteroids?

*Martin Elvis*

*Harvard-Smithsonian Center for Astrophysics*

**Abstract**

A simple formalism is presented to assess how many asteroids contain ore, i.e. commercially profitable material, and not merely a high concentration of a resource. I apply this formalism to two resource cases: platinum group metals (PGMs) and water. Assuming for now that only Ni-Fe asteroids are of interest for PGMs, then 1% of NEOs are rich in PGMs. The dearth of ultra-low delta-v (<4.5 km s$^{-1}$) NEOs larger than 100 m diameter reduces the ore-bearing fraction to only ~1 in 2000 NEOs. As 100 m diameter NEOs are needed to have a value ≥US$1 B and the population of near-Earth objects (NEOs) larger than 100 m diameter is ~20,000 (Mainzer et al. 2011) the total population of PGM ore-bearing NEOs is roughly 10. I stress that this is a conservative and highly uncertain value. For example, an order of magnitude increase in PGM ore-bearing NEOs occurs if delta-v can as large as 5.7 km s$^{-1}$. Water ore for utilization in space is likely to be found in ~1/1100 NEOs. NEOs as small as 18 m diameter can be water-ore-bodies because of the high richness of water (~20%) expected in ~25% of carbonaceous asteroids, bringing the number of water-ore-bearing NEOs to ~9000 out of the 10 million NEOs of this size. These small NEOs are, however, hard to find with present surveys. There will be ~18 water-ore-bearing NEOs >100 m diameter. These estimates are at present highly imprecise and sensitive to small changes, especially in the maximum delta-v allowed. Nonetheless the low values found here mean that much improved determinations of each of the terms of the formalism are urgently needed. If better estimates still find small numbers of ore-bearing NEOs then thorough surveys for NEA discovery and, especially, characterization are needed. Strategies for the two classes are likely to be different.

## 1. Introduction

For the mining of asteroids to become an engineering and commercial reality requires that we make a good assessment of how many asteroids contain ore. Here I use the term "ore" in the sense used in the terrestrial mining community, i.e. "*Ore is commercially profitable material*" (e.g. Sonter 1997). Ore is not simply a high concentration of some resource, but includes consideration of the cost of extraction of the resource and its price. Hence we need to sieve the total asteroid population for the smaller populations that may be profitable to mine. Main Belt asteroids are to hard to reach, so I will concentrate on the population of near-Earth objects (NEOs) which is overwhelmingly made up of asteroids, but with an admixture of comets. The NEO population is large. There are ~20,000 NEOs larger than 100 m diameter (Mainzer et al. 2011), and of order 10 million larger than 20 m diameter (Brown et al. 2013).

In this paper I introduce a simple formalism to evaluate how many ore-bearing asteroids there are. This formalism is likely to be reasonably robust. I then apply this formalism to two cases, the platinum group metals (PGMs) and water, using the limited available data. The values resulting from this analysis are by no means definitive, but the resulting values are quite small. The small numbers imply that further investigations to improve these estimates are urgent. Some possible paths forward are discussed for each term.

## 2. Quantifying the Question

We can quantify the number of ore-bearing NEOs, $N_{ore}$, for a given resource as the product of $P_{ore}$, the probability that an NEO is ore-bearing, and $N(>M_{min})$, the number of NEOs larger than a minimum profitable mass, $M_{min}$, for that resource:

$$N_{ore} = P_{ore} \times N(>M_{min}) \quad \text{(Equation 1)}$$

$P_{ore}$ is then the product of several factors[1]:

$$P_{ore} = P_{type} \times P_{rich} \times P_{acc} \times P_{eng} \quad \text{(Equation 2)}$$

Here $P_{type}$ is the probability that an asteroid is of the resource bearing type, $P_{rich}$ is the probability that this type of asteroid is sufficiently rich in the resource. The product of $P_{type}$ and $P_{rich}$ determines the fraction of NEOs with a high concentration.

In addition to a high resource concentration, $C_r$, qualifying an NEA as ore-bearing requires economical extraction of the resource, including its return to a location where it can be sold. I use two terms to quantify this challenge. $P_{acc}$ is the probability that the asteroid is in an accessible orbit and is discussed in Sec.3. $P_{eng}$ is the probability that the resource can be extracted profitably, as discussed in Sec.6. Other factors can be added to these equations as the calculations become more refined, but these capture the essence of the problem.

$N(>M_{min})$ depends the retrievable ore value in the asteroid, $\Lambda_{ore} = \varepsilon M C_r \lambda$, where $\varepsilon$ is the resource extraction efficiency which will likely be substantially less the unity, at least initially (Kargel 1994), and where $\lambda$ is the price/kg of the resource at the point where it can be sold, either on Earth or at various locations in space (see Sec.5). The total revenue must yield an acceptable profit given the substantial risk and long timescale of asteroid mining ventures.

Asteroid masses are hard to determine without sending a spacecraft close to the NEO. Only one mission, Hayabusa, has gone to a sub-km-sized NEO (Fujiwara et al. 2006). A minority of NEOs are binaries. For these, and for those undergoing close flybys of other massive bodies, Kepler's third law allows a mass to be derived (Merline et al. 2002). Radar can determine masses for NEOs passing close to Earth(≤ 0.1 AU, Ostro et al. 2002[2] ). But for the majority of NEOs a mass must be inferred from an assumed mean density and a diameter, so we must use a minimum diameter, $D_{min}$, as a proxy for $M_{min}$.

---

[1] This formalism is the same as that of the Drake equation for estimating the number of civilizations in the Galaxy capable of being detected (http://www.seti.org/drakeequation). Fortunately, the asteroid case has two fewer terms and better determined values.

[2] See also the plots at http://echo.jpl.nasa.gov/~lance/snr/far_asnr18.gif).

The resource extraction process includes a myriad of engineering details, which I subsume into $P_{eng}$. Evaluating $P_{eng}$ is too complex to include in this paper (see the discussions in Kargel 1994, Lewis et al. 1993). Hence I will take $P_{eng}$ = 1 throughout, so that all estimates of $N_{ore}$ given in this paper should be taken as upper limits. Some issues related to $P_{eng}$ are discussed in Sec. 6, including the possible dependence of the other terms on $P_{eng}$, which would spoil the simple factorization of equations 1, 2 by adding joint probabilities.

The numbers needed to evaluate equations 1 and 2 are at present mostly not well determined. Here I collect the available data in order to make an initial estimate of $P_{ore}$ and $N_{ore}$ for two much discussed cases – platinum group metals (PGMs) and water. The results are instructive. In the discussion I consider how to improve these estimates, how to increase $N_{ore}$, and how to find the ore-bearing NEOs.

## 3. Accessibility

Accessibility is primarily determined by the energy needed to go out to the asteroid with the mining equipment and to return with the ore. This energy is conventionally measured by delta-v, the change in velocity needed to transfer between orbits. The minimum energy trajectory is called a Hohmann transfer orbit (Hohmann 1960). The outbound delta-v can be approximated using the Shoemaker and Helin (1978) formalism. The return delta-v is more important than the outbound delta-v because a much larger mass of ore needs to be returned than the mass of the mining equipment sent out. Small changes in delta-v make for large differences in the mass that can reach an NEO (Elvis et al. 2011).

Benner has computed the outbound Hohmann delta-v values for all known NEOs from low Earth orbit (LEO) to an asteroid rendezvous orbit[3]. Values range for 3.8 km s$^{-1}$ to 28.0 km s$^{-1}$, with a median of 6.65 km s$^{-1}$ (Figure. 1, Elvis et al. 2011). Given the large payloads that mining missions, or a human expedition, would require, a lower delta-v is needed. Elvis et al. (2011) show that choosing an NEO with delta-v = 4.5 km s$^{-1}$ can double, or even quadruple, the payload delivered to the NEO compared with the median. This value includes only a small fraction of all known NEOs (Figure 1, black line).

The orbital dynamics that scatters asteroids into NEO orbits has no dependence on mass (Bottke et al. 2002), hence it is expected that size and orbit parameters are uncorrelated in the full NEO population. However, present surveys for NEOs are incomplete. In the 100 m – 300 m size range (roughly an absolute magnitude[4], H ~

---

[3] http://echo.jpl.nasa.gov/~lance/delta_v/delta_v.rendezvous.html.

[4] An asteroid's absolute magnitude H is the visual magnitude an observer would record if the asteroid were placed 1 Astronomical Unit (AU) away, and 1 AU from the Sun and at a zero phase angle (http://neo.jpl.nasa.gov/glossary/h.html). Conversion from H to an approximate diameter is given at http://neo.jpl.nasa.gov/glossary/h.html. H=22 corresponds to a diameter between 110 m and 240 m for typical albedos.

22) over 80% of NEOs remain undiscovered (Mainzer et al. 2011). The larger known NEOs with H < 22 have a higher median delta-v (8 km s$^{-1}$) than the smaller (H > 22) known NEO population (6.4 km s$^{-1}$). This is a selection effect in the known population as smaller NEOs can only be found when they are closer and so are more easily found if they have relatively Earth-like orbits.

For the H > 22 NEOs delta-v = 4.5 km s$^{-1}$ corresponds to $P_{acc}$ = 2.5% (Figure 1, blue line). To reach $P_{acc}$ = 25% requires only delta-v = 5.7 km s$^{-1}$, so $P_{acc}$ is highly sensitive to the choice of delta-v cut. Very small NEOs (24 – 60 meters, 25 < H < 27, Figure 1, green line) have a similar distribution to all H > 22 objects at low delta-v. Larger NEOs (diameter >100 m, H < 22) have $P_{acc}$ = 0.1% at delta-v = 4.5 km s$^{-1}$ and reach $P_{acc}$ = 10% only at delta-v = 6.2 km s$^{-1}$ (Figure 1, black line). As larger NEOs are more easily found their distribution should be more representative of the full NEO population. Both improved modeling of the NEO population (e.g. Greenstreet and Gladman 2012) and more complete observations can clearly have a big effect on our assessment of $P_{acc}$.

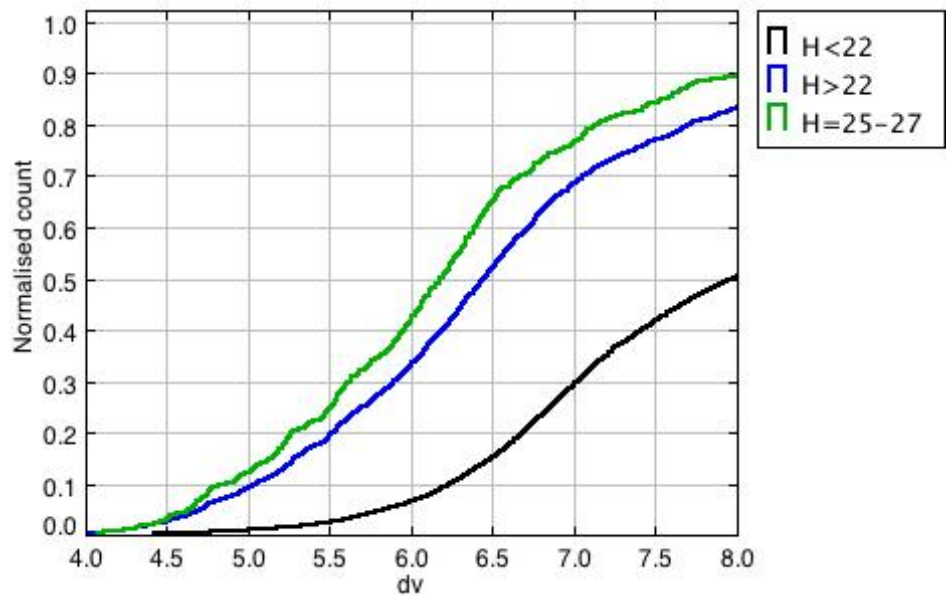

Figure 1. Cumulative distribution of outbound delta-v values for NEOs. Black curve: NEOs with H < 22 (diameters < 100 m); blue curve: NEOs with H > 22 (diameters >100 m); green curve: NEOS with 25 < H < 27 (diameters 24 – 60 meters, for low, 0.05, albedo objects).

## 4. Platinum Group Metals

First I consider the platinum group metals (PGMs): platinum (Pt), rhodium (Rh), osmium (Os), iridium (Ir), palladium (Pd), and rhenium (Re). These elements are rare in the Earth's crust as they are siderophiles – i.e. they dissolve readily in molten

iron – and so are mostly trapped in the Earth's core. As a result, several researchers (e.g. Kargel 1994) have identified the PGMs as the most promising asteroidal ore, because of their high value on Earth (~US$50k/kg, approximate present Pt prices[5]). Selling asteroid-derived resources on Earth has the advantage of not needing the development of a market for the resource in space.

Kargel (1994) discusses the fraction of NEOs that might be rich in PGMs, $P_{type}$, in Equation 1. The answer is not simple, as the choice of asteroid type depends on the ore extraction method. The richest asteroids would be M-type, which are thought to be those that deliver PGM-rich nickel-iron (Ni-Fe) meteorites to Earth, a subset of the X-class asteroids. I will concentrate on these, realizing that improved estimates for other PGM-rich asteroids need to be developed.

Binzel et al. (2004) compiled statistics for both Bus-Binzel (Bus et al. 2004) and Tholen (1984) classifications of NEOs. They find only three M-type NEOs. However, in addition, 27% of the 47 X-class asteroids, which have rather ambiguous spectra, will turn out to be M-type, if the ratios of E, P, and M types that make up the X-class remain the same (4, 4, 3, respectively). Adding these expected M-types gives a total of 16 out of a sample of 376. Hence $P_{type} \sim 4\%$. This percentage is comparable to that in meteorite falls (Grady 2000). The fraction in general meteorite collections is larger, both because they survive passage through the Earth's atmosphere much better than stony meteors (Sears 1998), and because they are more readily identified as meteoritic on the ground.

Distributions of PGM concentration, $C_r(PGM)$, for various meteorite groups are not widely available. Within the Ni-Fe meteorites the richness of Iridium (Ir) is better measured. Ir concentrations span four orders of magnitude, from 0.01 to 100 parts per million (ppm, Kargel 1994). Where both are measured, Iridium is well correlated with the other PGM content, though the number of meteorites analyzed is modest (Cook et al. 2004). The sum of all PGMs amounts to ~7 times the Ir richness in 90th percentile iron meteorites (Kargel 1994). Figure 2 shows the distribution of iridium concentration in ppm for 140 metalliferous type IIIAB meteorites. The data come from table 2 of Scott, Wasson and Buchwald (1973). (Newer papers report only mean values.) The curve is quite steep, with only a small tail of high concentration examples. Table 1 shows the mean richness of this data in 10% percentiles. Good terrestrial mines have PGM concentrations of up to 2-6 ppm, or grams per metric tonne (mt)[6]. The required Ir richness to match good terrestrial ore is then ~0.3 - 0.9 ppm. This concentration or higher is found in the top 50th percentile of type IIIAB meteorites (Table 1). So $P_{rich} \sim 50\%$.

The product $P_{type}$ x $P_{rich}$ = (0.04 x 0.5) tells us the fraction of NEOs having enough PGMs to consider mining. (I.e. 1/50 of all NEOs.) However, adding the tight constraint ($P_{acc}$ = 0.025) of a maximum delta-v = 4.5 km s$^{-1}$ from Section 3 (and assuming $P_{eng}$ =1) gives the final probability of a NEO being PGM ore-bearing,

$$P_{ore}(PGM) = P_{type} \text{ x } P_{rich} \text{ x } P_{acc}(4.5 \text{km s}^{-1})$$

---

[5] http://www.platinum.matthey.com
[6] http://www.platinum.matthey.com/about-pgm/production/south-africa

$$= 0.04 \times 0.5 \times 0.025$$
$$= 5.0 \times 10^{-4}.$$

That is 1/2000 of the total NEO population.

Table 1: Mean resource concentration (ppm) in 10% percentile ranges for type IIIAB iron meteorites. Total PGM richness is set to be 7 times higher than Ir (Kargel 1994).

|            | Mean resource concentration (ppm) ||
| ---------- | ------ | -------- |
| Percentile | Ir     | All PGMs |
| 90         | 11     | 75       |
| 80         | 5.6    | 39       |
| 70         | 3.9    | 27       |
| 60         | 2.1    | 15       |
| 50         | 0.97   | 6.8      |
| 40         | 0.50   | 3.5      |
| 30         | 0.26   | 1.8      |
| 20         | 0.068  | 0.47     |
| 10         | 0.024  | 0.17     |

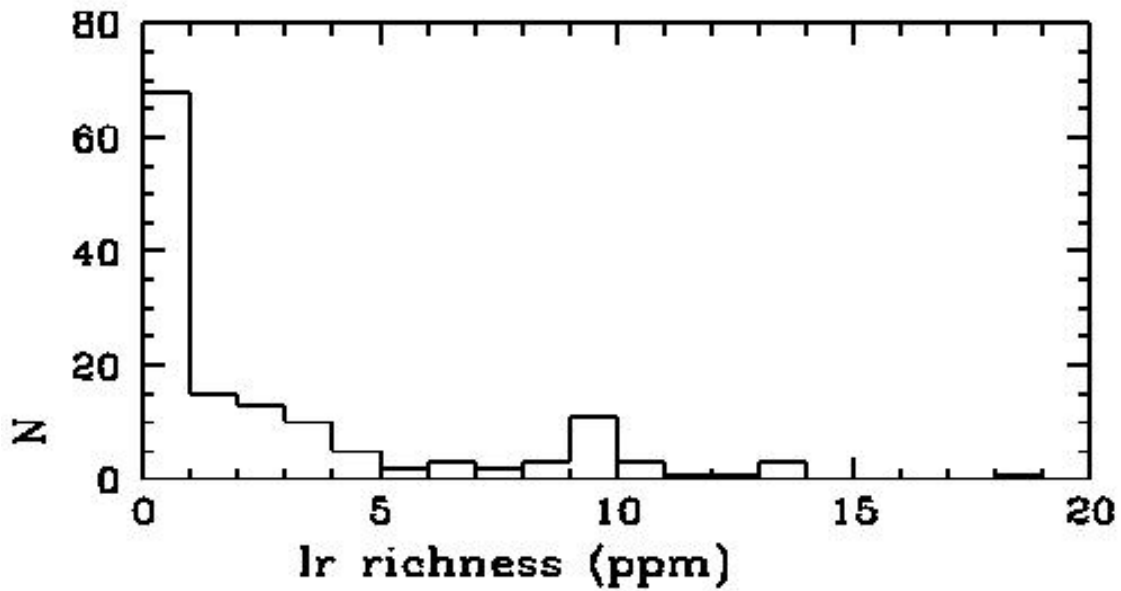

Figure 2. Distribution of Iridium (Ir) richness (ppm) in metalliferous type IIIAB meteorites (data from table 2 of Scott, Wasson and Buchwald, 1973). 50% have Ir richness >1ppm, corresponding to a total PGM richness >7 ppm, comparable with the best terrestrial Pt mines.

Table 2. Parameter Estimates for PGM- and Water-ore bearing NEOs.

| Parameter | PGMs | Water |
|---|---|---|
| $P_{type}$ | 4% | 25 - 50% |
| $P_{rich}$ | 50% | 25% |
| $C_r$ | 10 ppm | 20% |
| $P_{acc}$[a] | 2.5% | 3% |
| $P_{ore}$[a] | $5 \times 10^{-4}$ | $8 \times 10^{-4}$ |
| $D_{min}$ (H) | 100 m (~22) | 18 m (~27) |
| M(NEO) (at $D_{min}$) | $2.4 \times 10^6$ mt | 1180 mt |
| $M_{ore}$ (at $D_{min}$) | 23.6 mt | 236 mt |
| Ore volume (at $D_{min}$) | 1.1 m$^3$ | 236 m$^3$ |
| Assumed Price, $\lambda$ $\Lambda_{ore}$ (at $D_{min}$) | US$50k/kg, US$1.2B | US$5k/kg, US$1.2B (at LEO[b]) US$10k/kg, US$2.4B (at GEO[c]) |
| N(>$D_{min}$) | 20,000 | 10,000,000 |
| $N_{ore}$[#] | 10 | 8000 18 (at H=22) |

a. for maximum outbound delta-v=4.5 km s$^{-1}$; b. Low Earth Orbit; c. Geostationary orbit.

What is $M_{min}$, the minimum mass NEO worth mining for PGMs? We measure diameter, not mass, telescopically, so we need a density to find $M_{min}$. Solid Ni-Fe has a density of 7,300-7,700 kg m$^{-3}$, but meteorites have both microporosity (on scales of a micron or smaller) and macroporosity (on larger scales), both of which lower the NEO density (Britt et al 2002). Small asteroids can have large macroporosity due to large cracks, fractures, or to being rubble piles. Carry (2012) lists only 17 NEO mass measurements, all but one of which come from radar, but none are M-type. For X-type asteroids Carry (2012) gives only six density measurements for moderately small (<50 km diameter) asteroids that have reasonable accuracies (50% or better). These have a mean density of 4500 kg m$^{-3}$. Recognizing the limitations of this data, including that only M-class asteroids, a subset of X-type, have high PGM content, I will use this value.

At this density a 100 m diameter asteroid would have a mass of 2.36 x10$^6$ metric tons (mt). A type IIIAB meteorite chosen at random from the top 50% has an order of magnitude range of $C_r$(PGM) from 6.8 to over 75 ppm (Table 1). If we conservatively take $C_r$(PGM) = 10 ppm then the mass of PGMs would then be 23.6 mt. (The 21.45 g cm$^{-3}$ density of platinum means that this mass occupies just 1.1 m$^3$.) If this mass of PGMs could all be extracted and returned to Earth then, at $\lambda$ = US$50 k kg$^{-1}$, $\Lambda_{ore}$ = US$1.18 B. Given the billion dollar costs of most interplanetary missions, the target must have $\Lambda_{ore} \geq$ US$1 B of PGMs, making $D_{min}$ = 100 m a plausible lower limit to the size of NEO worth mining. The range of $C_r$(PGM) means that $\Lambda_{ore}$ ranges from a low of US$0.8 B to a high of US$8.8 B. Such a large range of values could greatly change the profitability of a venture, making more accurate assays necessary.

Note that I am assuming here that asteroid mining does not flood the market and depress PGM prices, which is plausible for the first deliveries as world platinum production is ~200 mt year$^{-1}$ (Loferski 2012), some ten times larger than one good 100 m asteroid would provide. Markets to not always respond linearly to changes in supply, however.

Larger asteroids quickly become more resource rich. An otherwise identical 160 m dia. NEO would have $\Lambda_{ore}$ = US$5 B of PGMs at 10 ppm. Because the slope, $\alpha$, of the size-frequency distribution of NEOs, $N \propto D^{-\alpha}$, is < 3, large NEOs will dominate the total value of PGM ore-bearing NEOs. As larger NEOs are rarer, knowledge of one will be valuable information.

Smaller asteroids become unpromising just as rapidly. A 50 m diameter asteroid would contain one eighth of the platinum mass (2.9 mt) and $\Lambda_{ore}$ = US$150 M. Good size and mass estimates are thus crucial to asteroid mining.

Resource extraction efficiency ($\epsilon$ in Sec.2) is unlikely to be 100%, especially at first (Kargel 1994). That may lead to a requirement for higher $C_r$. If we increase $C_r$(PGM) to five times that of terrestrial mines then $P_{rich}$ = 0.25, only a factor two reduction.

Given the, as yet unknown but likely substantial, cost involved in a mining mission, $D_{min}$ = 100 m seems like an optimistic estimate of the threshold for profitability. N(D>100), the number of NEOs larger than 100 m diameter, is about 20,000 (Mainzer et al 2011).

Combining these estimates the total number of PGM ore-bearing asteroids, $N_{ore}$(PGM), is likely to be about 10 (Table 2). I stress that this number has large uncertainties and includes only metallic asteroids. Nonetheless, the number is surely smaller than would-be asteroid miners may have expected.

Pallasite meteorites probably have the same parent body as the IIIAB irons (Tarduno et al. 2012). They are the next best candidates for PGM mining. The ~50 known examples have a mean 50% total iron content by weight and span only a narrow range in iron content (38% - 70%, Buseck 1977). Of the 33 pallasites tabulated by Wasson and Choi (2003), three (9%) have greater than 10 ppm of Pt (Cold Bay at 20 ppm, Eagle Station at 22 ppm, Yamato 8451 at 17 ppm). They are also rich in Re and/or Ir. Two more (6%) are rich only in Re, from 235 to 1235 ppm. These 20 – 120 times higher values of $C_r$ more than compensate for the 10 times lower price of Re at around $\lambda$ = US$5k/kg[7]. So $P_{rich}$ = 0.15. Pallasites are strongly associated with A-type asteroids (Lucey et al. 1998). Binzel et al. (2004) find only

---

[7] www.roskill.com/reports/minor-and-light-metals/rhenium, 2013 January 31.

one A-type NEO out of 376, so $P_{type}$ = 0.0027, a negligible addition to the M-class numbers. As A-type NEOs have distinctive spectra (they are end members of a principal component, DeMeo et al. 2009) they are readily identified and there is no pool of ambiguous type NEOs from which to increase their numbers. For now, therefore, I ignore this group.

The strongest limitation on $P_{ore}$ for PGMs is delta-v. If the allowed outbound delta-v can be raised by just over 1 km s$^{-1}$, to 5.7 km s$^{-1}$, then $P_{acc}$ = 25% (Sec. 3) then the number of PGM-ore-bearing NEOs will be an order of magnitude larger, ~100.

**5. Water**

Water is often considered the first product likely to be mined from space. The water would be used in space either for life support or, separated into hydrogen and oxygen, for rocket fuel. Water, or any other resource mined for use in space, is valuable because it can substitute for material that would otherwise be brought up from Earth. Historically, to take a kilogram of anything from the ground to low Earth orbit (LEO, at an altitude of ~200 – 2000 km) has cost ~US$10 k or more for decades (in 2000 dollars, Futron Corporation 2002).

Water supplied from space would have to undercut the cost from Earth, say by a factor two. The mined water could then be sold for λ~US$5k/kg, about a factor 10 less valuable in LEO than PGMs are on Earth. The value of water doubles in geostationary orbit (GEO, altitude 35,786 km), because of the extra energy needed to reach this higher orbit from Earth, and the rocket equation, which calculates the penalty of having to carry propellant with the rocket. This penalty is even higher at the Earth-Moon L1 Lagrange point (~340,000 km from Earth), so that the cost of bringing water there from Earth perhaps equals the λ of PGMs. For now LEO seems to be the most likely location for the first sales of water, so I will calculate $P_{ore}$ using the lower λ value applicable there.

The size of the market for water in space is presently small. For a crew of 3 in the year from May 2011 to May 2012 3.3 mt were delivered to the International Space Station (ISS) (J.C. McDowell, private communication) for a total cost of ~US$33 M/year. The in-space market for asteroidal water would lose value if lower launch costs/kg emerge. However, the existence of a supply of cheap water in space may itself stimulate the growth of a market.

If the goal is to mine water, then the larger population of carbonaceous (C-type) asteroids is the most likely source of water. C-types comprise (9.8±3.3)% of the NEO population, correcting for bias due to their low albedo (Stuart & Binzel 2004). Hence $P_{type}$ = 0.1 is a reasonable value.

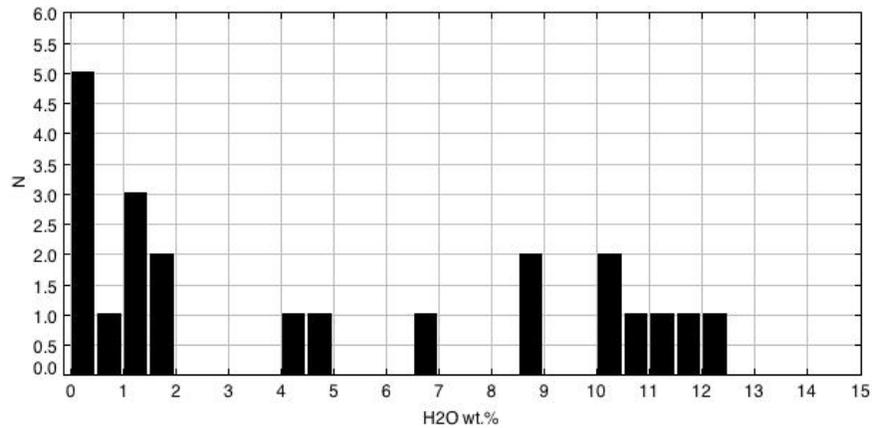

Figure 3: Distribution of H$_2$O by percentage by weight (wt.%) for Carbonaceous chondrites (Jarosewich 1990).

Asteroidal water is either bound up in hydrated minerals such as clays, or may be present as ice distributed among the rock.

Ice can only exist under a surface layer that protects the ice from sublimation. The size of the thermal skin depth might impose a minimum size for ice-bearing asteroids, although Harris and Lagerros (2002) give thermal skin depth values of millimeters to centimeters. The discovery of briny fluid inclusions in some meteorites (Saylor et al. 2001) supports the existence of ice in asteroids. The curation of the Antarctic Search for Meteorites (ANSMET) program involves thawing the meteorites to remove ice (Allen et al. 2013), which may have inadvertently led to low estimates of their ice content. Extraction of water from hydrated minerals has been considered (Lewis 1996) and found plausible.

The amount of ice in an NEO would depend on the porosity of the rock. (I.e. the fraction of the body that is "empty" space.) Carbonaceous meteorites have microporosities of 1% to ~20% (Britt and Consolmagno 2003), which gives an upper limit to their small scale ice content. Distributions of microporosity for carbonaceous meteorites are not available. Macroporosity is measured only for three C-type NEOs in Carry (2012) at 28±37, 39±47 and 60±32 percent (3671 Dionysus, 1996 FG3 and 2002 CE26, respectively). These porosities allow commensurately higher ice content with, obviously, high uncertainty.

The fraction of water bound into silicate clays is better measured in carbonaceous chondrites than the ice content. A distribution of the percentage water fraction by weight is given for 22 meteorite samples by Jarosewich (1990) and shown in figure 3. The lowest bin contains three upper limits at 0.1 wt.%. There appears to be a group of high H$_2$O concentration (> 6 wt.%) containing 28% of the points (9/22). However of the eight points with >6 wt.% Jarosewich notes that the three Murchison samples are not independent samples, and the two ALH meteorites also appear to have a common origin. Allowing for these 31% (6/19) is a better estimate of the high water concentration group $C_r$(ice).

From the hydrated mineral distributions I take $P_{rich}$ =0.25. The fraction of water-rich NEOs is then $P_{type}$ x $P_{rich}$ = 0.025.

For smaller (H > 22) NEOs $P_{acc}$ = 3%, based on an outbound delta-v = 4.5 km s$^{-1}$ (Sec. 3, Figure 1, green curve). Hence the probability of a NEO being water-ore-bearing (table 2) is:

$$P_{ore}(water) = P_{type} \times P_{rich} \times P_{acc} = 0.1 \times 0.31 \times 0.03 = 9 \times 10^{-4}$$

(As for the PGM case I do not consider $P_{eng}$.) Roughly 1/1100 NEOs are thus water-ore-bearing, quite comparable to PGM ore-bearing asteroids.

However, the 20,000 times higher $C_r$ for water-rich NEOs compared with PGM-rich NEOs opens up the much more numerous population of smaller NEOs, even for delivery to LEO. Given the water concentration of carbonaceous chondrites from Jarosewich (1990) I use a $C_r$(hydrated) = 10% plus $C_r$(ice) = 10% for a total $C_r$(water) = 20%. From this we can estimate the value of the resource and so the number of resource rich NEOs. If mission costs do not depend strongly on whether PGMs or water is being mined, then the threshold for profitability should be the same. To reach $\Lambda_{ore}$ = US$1.18 B, as used for PGMs (Sec. 4), with a $C_r$(water) = 20% requires only the water from an 18 m diameter carbonaceous NEO delivered to LEO and sold at $5 k kg$^{-1}$. A 30 m diameter NEO has $\Lambda_{ore}$ = $5 B, a more likely to be profitable value.

A recent redetermination of the number of 20 m class impacts on the Earth's atmosphere, calibrated using the Chelyabinsk event, has increased the estimate of the number 20 m or larger NEO by a factor ~10 to about ~10 million (Brown et al. 2013). As the NEO frequency-size curve is steep between 20 m and 100 m, this new estimate is sensitive to small changes in the required $D_{min}$. With $P_{ore}$ = 0.0009 for these small asteroids, the total population of water-ore-bearing NEOs, $N_{ore}$(water) ~9000. This is over two orders of magnitude more that for PGM-ore-bearing NEOs. As with PGMs, a small increase in delta-v to 5.5 km s$^{-1}$ would increase $P_{acc}$ and so $N_{ore}$(water) by an order of magnitude to ~90,000.

NEOs of this smaller size are hard to discover and characterize. The absolute magnitude of an 18 m diameter asteroid is H = 26 – 27.5, depending on albedo (see footnote 2). The C-type asteroids tend to have low albedo (~0.01 - ~0.2, Stuart and Binzel 2004, Ryan and Woodward 2010, Thomas et al. 2011), so H = 27 is more appropriate. To be detected by the main current asteroid surveys an asteroid must be at least as bright as V = 20. To be that bright an H = 27 NEO must come within 15.4 lunar distances (= 5.9 x 10$^6$ km = 0.04 AU). The discovery rate for NEOs with H > 25 is currently ~250/year, based on Minor Planet Center statistics (Beeson et al. 2013). Since $P_{ore}$(water) = 1/1100, roughly one water-ore-bearing NEOs every five years is now being found. For the most part, we do not know which ones these are, as follow-up colors and spectra for these small objects are sparse. Moreover, because most remain detectable for only a few days, most have poorly determined orbits (with orbit uncertainty parameters[8], U > 5), and so are effectively lost (Beeson et al. 2013).

If instead we use the number of the more readily found H < 22 NEOs then the steep frequency-size curve for NEOs in this size range gives $N(>D_{min})$ = 20,000, as for the PGM case, and $N_{ore}$(water) ~ 18. Clearly improved surveys to find and characterize small NEOs would be extremely helpful in making the profitable mining of asteroidal water feasible.

**6. Engineering Challenges, $P_{eng}$**

The term $P_{eng}$ is included in equation 2 as a way to capture all the uncertainties involved in extraction of the sought resource. I do not attempt to evaluate these factors here, but list some examples of the issues to be addressed.

A "tumbling" (i.e. non-principal axis rotator) asteroid will be more demanding to anchor a miner spacecraft to than a simple principal axis rotator. In this case it is not possible to achieve a smooth low energy anchoring procedure by putting the spacecraft at the pole and rotating it at the asteroid period. Instead, continuous active thrusting will be required until anchoring is achieved. This is a potentially costly activity in energetic terms.

Solid rock or Ni-Fe demands large energy inputs, by space mission standards, to extract PGMs, ~10 MW for 20 mt/year (Kargel 1994). For comparison, a 40 kW solar electric propulsion system is now considered high power (Brophy et al. 2010), a factor 250 less. A year is a plausible dwell time at the asteroid for both orbital mechanics and financing reasons. Extraction of the PGMs would likely occur at the asteroid, as returning an intact 2 million metric tonne asteroid to cis-lunar space is both a daunting propulsion challenge (though less so if 10 MW were available) and would raise obvious safety concerns.

Water should be operationally easier to mine, reducing mission costs. For example, processing water probably requires much less energy (Lewis 1996), and the complications of re-entry through Earth's atmosphere are removed as the resource is to be sold and used in space. The high resource concentration for water makes return of the entire asteroid a plausible option (Table 2), so extraction could occur in cis-lunar space. This allows more massive processing facilities and their re-use.

On the negative side, water purification to levels adequate for use as either use as fuel or for human consumption may be complex. Carbonaceous asteroids are thought to be rich in complex organic molecules and these could cause problems in both use cases. For use as fuel the resulting hydrogen and oxygen would need to be liquefied and stored at cyrogenic temperatures. Water also requires larger masses and much larger volumes to be returned than PGMs (Table 2) in order to be profitable.

Moreover, in all cases, resource extraction efficiency ($\varepsilon$ in Sec.2) is not likely to approach 100%.

---

[8] See definition at: http://www.minorplanetcenter.net/iau/info/UValue.html

These challenges are not necessarily simply factorizable to fit the schema of equation 2, but can have complex interconnections. For example, the richness that can be mined may well depend on engineering details. In this case the two cannot be considered separately. We can see the beginning of that in the dependence of $D_{min}$ on $P_{rich}$ (Sec. 4). Nonetheless, the $P_{eng}$ term cannot be neglected in commercial evaluation of whether or not an asteroid is ore-bearing.

Without more analysis of $P_{eng}$ we cannot know if mining of an NEO would be profitable.

## 7. Discussion

The initial estimates of $N_{ore}$ presented here suggest that it is likely that there are relatively few ore-bearing NEOs. It is the need to fulfill three criteria simultaneously that creates these small numbers. Recall also that I have unrealistically set $P_{eng} = 1$. I estimate that PGM-ore-bearing NEOs (with diameters >100 m) number a handful, while water-ore-bearing NEOs (with diameters >18 m) number a few thousands. However, most of these latter small NEOs will be hard to find and characterize. The score or so water-ore-bearing NEOs larger than 100 m diameter would be easier objects to find.

The most important conclusion is that this formalism usefully exposes the key factors for asteroids to be ore-bearing, and that examining them shows that all the values for $P_{type}$, $P_{rich}$, $P_{acc}$ and $N(>M_{min})$ used to make this assessment are in need of far better definition. Some of these asteroid prospecting considerations are given in Elvis (2013). Below I discuss how to improve the estimates for each factor in turn. They are all significant research projects.

**$P_{acc}$** can be improved with calculation of outbound and return delta-v values for real trajectories rather than Hohmann transfer orbits. These trajectories are time-dependent and so are computationally intensive. The Near-Earth Object Human Space Flight Accessible Targets Study (NHATS[9]) calculated outbound and return delta-v for several thousand known NEOs every 8 days over a 30 year interval using large amounts of compute time on a major supercomputer. To do this for a complete simulated NEO population would be instructive. Alternative delivery orbits should also be considered, including LEO, GEO and the Earth-Moon L1 Lagrange point.

Modest increases in the maximum accessible delta-v will raise $P_{acc}$ by a substantial factor for both PGMs and water. delta-v would have to be 6.5 km s$^{-1}$ to increase $P_{acc}$(PGM) to 50% for small (H > 22) NEOs (Figure 1). Investment in better propulsion systems will raise $P_{acc}$. High power solar electric propulsion seems to be the currently most straightforward way to increase the acceptable delta-v for NEO missions (Brophy et al. 2010), though nuclear propulsion options are promising in principle (Houts et al. 2012).

---

[9] http://neo.jpl.nasa.gov/nhats/

Novel return trajectories exist for a fraction of the NEO population with an order of magnitude lower return delta-v (Sanchez and McInnes 2011, Garciá Yárnoz et al. 2013) over a wider range of orbital parameters than low delta-v Hohmann transfer orbits (Elvis et al. 2011).

**$P_{type}$** is poorly known for most NEOs. Spectroscopic characterization is proceeding at ~100/year (R.P. Binzel, private communication), well behind the discovery rates. At least an order-of-magnitude increase in characterization rates for NEOs is needed to find useful numbers of ore-bearing NEOs. This is feasible at modest cost if only optical spectra are needed (Elvis et al. 2013). A better estimate of the parent population of meteorites could be found from Antarctic meteorites, as stony and carbonaceous meteorites are as visible on the snowpack as the iron meteorites, which is not the case in most other locations. Curation that avoids loss of asteroidal ice is desirable.

**$P_{rich}$** is not generally determinable using remote, telescopic, methods. However, all of the water rich (> 8%) meteorite samples in Jarosewich (1990) are CM and CI chondrites. CM chondrites can be identified with good optical-near-IR spectra as Ch-type ("Carbonaceous hydrated") asteroids from the 0.7 micron and 3.1 micron hydration absorption features (Burbine 1998, Bus et al. 2000, Rivkin et al. 2002). The easily measured 0.7 micron feature is weak in NEOs (Howell et al. 2011), leaving the technically demanding 3.1 micron feature as the best test. A large campaign of NEA spectroscopy might then raise $P_{rich}$ close to 1 for water. Other types of meteorites should be included to assess $P_{rich}$(water). Radar measurements of circular polarization may efficiently select metallic asteroids as they seem to have smooth dielectric surfaces (Benner et al. 2008).

In most cases, in order to perform a detailed assay of resource content asteroid prospecting needs to go beyond astronomy-based techniques. Improved measurements of large numbers of meteorites of all types would define the distributions far better than at present (Figures 2, 3). Connecting meteorites to classes of asteroids is difficult. A deliberate search for small asteroids about to impact the Earth would be immensely valuable. The only instance to date is 2008 TC3 (Jenniskens et al. 2009), but objects of this size impact the Earth roughly every month (Brown et al. 2002).

**$N(>M_{min})$** is limited in accuracy by the paucity of good estimates of the masses of potential target asteroids. Greatly expanded numbers are clearly needed. Radar techniques provide the best measurements, but cannot yet reach beyond ~0.1 AU (see Sec. 2).

Even with all the possible improvements, in order to bring down the risk for a specific mining expedition to a prudent level it will be necessary to make measurements local to the NEO, but only for high priority targets, before mining operations begin. The number of assay probes needed is addressed in Elvis and Espy (2013, submitted).

**P_eng** is in great need of detailed study. Studies with asteroid simulants on the International Space Station could begin to define $P_{eng}$. Some of the studies being undertaken for the Asteroid Redirect Mission (ARM[10], see Brophy et al. 2010), including the issues involved in capture (Roithmayr 2013), will also begin to address $P_{eng}$. If a small, ~7 m diameter ~500 mt, asteroid were returned to cis-lunar space, as envisaged for ARM, it could be used to test asteroid mining techniques leading to improved $P_{eng}$.

Equations 1 and 2 assume that the different terms are simply factorizable, i.e. each one is independent of the others. In reality there will be joint probabilities. If, for example, $P_{eng}$ implies a lower retrieval rate or a higher cost, then the amount of ore returned must be higher, increasing $M_{min}$, and this in turn reduces $N_{ore}$, as there are fewer large asteroids.

## 8. Conclusions

I have presented a simple formalism to assess the number of ore-bearing near-Earth objects. The most important conclusion of this study is that this formalism exposes the key factors for asteroids to be ore-bearing, and that examining them shows that all the values for $P_{type}$, $P_{rich}$, $P_{acc}$ and $N(>M_{min})$ used to make this assessment are in need of far better definition.

Initial estimates give very low values for platinum group metals, and larger, but still modest, numbers for water. The formalism of equations 1 and 2 is fairly robust, though it is subject to joint probabilities (Sections 6, 7). However, the values for $P_{ore}$ given here are just initial estimates to see where the numbers appear to lie, and to identify the places where improvement is most needed, as discussed in Sec. 7. Significant research is needed in all areas: on meteorite composition, on telescopic discovery and characterization, and on both of these together for the small asteroids about to impact the Earth, on local assay probes, and on in-space engineering challenges, both at the ISS and with a returned small asteroid.

The apparently limited supply of potentially profitable NEOs argues strongly for an accelerated rate both for discovery and especially for characterization, which is lagging badly behind discovery. At the expected discovery rate of about 2000 NEOs/year from 2015 onwards, and ~800 NEOs/year with H < 22, it will take a decade or so to essentially complete the discovery surveys (Beeson et al., 2013). Improved characterization surveys, which currently run at ~100 NEOs/year, are clearly needed for NEO orbits, size and compositions. These surveys should reach down to a 100 m scale (H ~ 22) for PGMs, and down to an 18 m scale (H ~ 27) for water. Different strategies may well apply to the two cases. In order to have at least one PGM-ore-bearing NEO requires a minimum completeness of nearly 100%. In contrast, a completeness of 10% for water-ore-bearing NEOs would yield ~900 objects, but must be more ~100 times (5 magnitudes) more sensitive.

As good targets appear to be scarce, the knowledge of which NEOs are ore-bearing could itself become commercially valuable intellectual property.

---

[10] http://targetneo.jhuapl.edu/pdfs/agenda.pdf

I thank Charlie Beeson, BC Crandall, Francesca deMeo, Doug Finkbeiner, José Luis Galache, Glenn MacPherson, Jonathan C. McDowell and Mark Sonter for valuable inputs. I also thank the two anonymous referees for forcing me to sharpen the arguments in the paper. The Aspen Center for Physics provided the essential peaceful environment that enabled me to answer the referees' careful points.